# Breakdown of Magic Numbers in Spherical Confinement

Junwei Wang[1,4]*, Jonathan Martín-González[1], Lukas Römling[1], Silvan Englisch[3], Chrameh Fru Mbah[2], Praveen Bommineni[2], Erdmann Spiecker[3], Michael Engel[2]*, Nicolas Vogel[1]*

[1]*Institute of Particle Technology, Friedrich-Alexander-Universität Erlangen-Nürnberg, 91058 Erlangen, Germany*

[2]*Institute for Multiscale Simulation, Friedrich-Alexander-Universität Erlangen-Nürnberg, 91058 Erlangen, Germany*

[3]*Institute of Micro- and Nanostructure Research and Center for Nanoanalysis and Electron Microscopy, Friedrich-Alexander-Universität Erlangen-Nürnberg, 91058, Erlangen, Germany*

[4]*Max Planck Institute of Colloids and Interfaces, 14476, Potsdam, Germany*

***Emails:** junwei.wang@mpikg.mpg.de, michael.engel@fau.de, nicolas.vogel@fau.de

## Abstract

Magic numbers in finite particle systems correspond to specific system sizes that allow configurations with low free energy, often exhibiting closed surface shells to maximize the number of nearest neighbors. Since their discovery in atomic nuclei, magic numbers have been essential for understanding the number-structure-property relationship in finite clusters across different scales. However, as system size increases, the significance of magic numbers diminishes, and the precise system size at which magic number phenomena disappear remains uncertain. In this study, we investigate colloidal clusters formed through confined self-assembly. Small magic number clusters display icosahedral symmetry with closed surface shells, corresponding to pronounced free energy minima. Our findings reveal that beyond a critical system size, closed surface shells disappear, and free energy minima become less pronounced. Instead, we observe a distinct type of colloidal cluster, termed football cluster, which retains icosahedral symmetry but features lower-coordinated facets disconnected by terraces. A sphere packing model demonstrates that forming closed surface shells becomes impossible beyond a critical system size, explaining the breakdown of magic numbers in large confined systems.

## Introduction

The magic number effect is a unique phenomenon observed in finite particle systems, predicting system size-dependent thermodynamic properties. Specific particle configurations, known as magic number clusters, exhibit exceptional stability, indicated by local free energy minima. These clusters typically have closed surface shells to maximize the number of nearest neighbors.[1,2] The magic number phenomenon was first observed in the unusual stability of certain nuclei against decay, where protons and neutrons arrange into closed shells.[3–5] Furthermore, magic number atomic clusters explain spikes of high occurrence in the mass spectra of Xenon clusters nucleated in the gas phase.[6] Since these initial discoveries, magic number atomic clusters have been observed in many chemical elements, displaying a remarkable variety of symmetries and structures.[2,7–10]

The magic number effect results from a competition between crystallinity and surface energy minimization and is directly related to the geometry of sphere packings.[1,2] For example, in a small droplet of condensing atoms, minimization of surface energy requires exposing low-energy facets. The facets with the lowest energy in the face-centered cubic (fcc) crystal are {111} facets, which form a hexagonal lattice with high coordination number.[11] To maintain crystallinity while exposing only low-energy facets, clusters with particularly low energy adopt a polyhedral Wulff shape with atoms arranged into concentric closed shells.[1] The most spherical polyhedron with only {111} facets is an icosahedron. The smallest icosahedral cluster consists of 13 atoms, with 12 atoms forming the first shell around one central atom. By successively adding more shells, a series of icosahedral clusters, known as Mackay clusters, can be constructed.[12,13] Removing or adding atoms to a Mackay cluster reduces its stability because it creates lower-coordinated surface atoms, rendering Mackay clusters local energy minima.

At a larger scale, colloidal particles self-assembled within the confinement of an emulsion droplet can also form magic number clusters as minimum free energy structures.[14,15] Colloidal clusters exhibit icosahedral symmetry with Mackay shells in the interior and additional anti-Mackay surface shells. The outermost colloidal particles form a characteristic surface shell structure, achieving a closed-shell configuration.[14,15] Magic number clusters with up to about 10,000 colloidal particles, differing by their number of Mackay and anti-Mackay shells, have been observed in experiments and simulations.[14] Compared to atomic clusters, the complex interaction potential of atoms is simplified to a hard sphere-like interaction for colloidal particles, with thermodynamic stabilization dominated by entropy.[16] Analogous to enthalpy minimization by forming closed surface shells with highly coordinated facets in icosahedral Wulff shapes, icosahedral colloidal clusters maximize entropy by improving packing efficiency within spherical confinement, resulting in a high coverage of the cluster surface with {111} facets.[14,16,17]

With increasing system size, colloidal clusters alternate between magic number regions, where the number of particles allows for closed surface shells, and off-magic number regions, where the formation of such closed surface shells is disrupted by the accumulation of defects.[15] Transitions between these regions are reflected by free energy fluctuations, with magic number configurations appearing as periodic minima.[18]

Magic number effects have been observed in diverse systems across length scales, from subatomic nuclei to atomic and molecular clusters, assemblies of proteins, nanoparticles, colloids, and even millimeter-sized objects.[2,4,19–30] However, despite the generality of magic numbers, two fundamental questions remain unanswered: (i) The observation of magic number configurations with up to 20,000 building blocks[31,32] exceeds Alan Mackay's expectation that geometric strain would render large icosahedral clusters unfavorable.[12] Is there an upper size limit above which the existence of magic number configurations breaks down? (ii) As system size increases, the discrete nature of finite systems eventually transitions into continuous, bulk behavior. Is the breakdown of magic numbers directly associated with this transition or do some characteristics of finite systems persist even when magic number configurations are no longer observed?

In this combined experimental and computational work, we use colloidal clusters as a model system to investigate the structural evolution from magic number clusters to bulk behavior. We analyze colloidal clusters containing from a few hundred to 400,000 particles. Our findings reveal that closed-shell clusters associated with magic number configurations disappear beyond a critical cluster size. This transition coincides with the attenuation of free energy minima. A sphere packing argument demonstrates that forming closed surface shells becomes geometrically impossible under spherical confinement once the system reaches the critical size. Beyond this point, the system does not directly revert to fcc clusters but instead forms a structurally distinct type of icosahedral cluster, which we term football clusters. Football clusters retain the icosahedral symmetry of smaller clusters but exhibit more lower-coordinated facets disconnected by terraces, leading to a more complex surface structure. Their appearance marks a significant shift in the structural organization of colloidal clusters and highlights the intricate balance between geometry, surface energy, and entropy in determining the stability and configuration of large finite systems.

## Results and Discussion

### Icosahedral Clusters without Closed Surface Shells – A Distinct Type of Colloidal Clusters

We fabricate emulsion droplets of an aqueous colloidal dispersion containing charge-stabilized polystyrene particles (PS) with a diameter of $\sigma$ = 244 nm in a perfluoroether (HFE 7500, 3M) continuous phase. We create these droplets either using droplet-based microfluidics or by vigorous shaking and subsequent emulsification into water-in-oil-in-water double emulsions.[33–35] The colloidal particles within the droplets are crystallized by gradual reduction of the confinement volume through osmotic pressure increase in the outer continuous phase. This equilibrates the colloidal particles into minimum free energy colloidal clusters.[14–16,35–38] We visually distinguish four types of colloidal clusters by their surface shell structure.[14,15]

Of particular importance is the presence of a closed surface shell. We call a surface shell *closed* if its particles are connected into a two-dimensional network by nearest-neighbor bonds that encompasses the entire cluster. {111} facets and {100} facets can be part of a closed surface shell because their particles are connected into a hexagonal lattice and a square lattice, respectively. In contrast, {110} facets cannot be part of a closed surface shell because their particles are only connected by nearest-neighbor bonds along a single direction.

We identify four types of colloidal clusters via electron microscopy as follows:

**Mackay Clusters:** These icosahedral clusters are common minimum free energy structures for small systems.[13,14,16] Figure 1a shows a Mackay cluster composed of 2,000 colloidal particles. Mackay clusters have 20 hexagonal {111} facets surrounding twelve five-fold symmetry axes. The {111} facets share edges (inset of Figure 1a).[13,14,16] A icosahedral sphere packing model accurately describes the observed surface shell (Figure 1a, with details in Figures S1-S2) and was validated through simulations and electron tomography studies.[14,15,39,40] The interior of a Mackay cluster consists of twinned tetrahedral fcc grains,[13,14] as illustrated by a single grain in Figure 1a, bottom left. In the five-fold vertices where five grain boundaries meet, no high-coordinated surface can be formed under the spherical confinement. Nearest neighbors of particles in this region constitute three-dimensional network of bonds, allowing defects to accumulate locally to reduce associated energy penalty.[15] However, Mackay clusters are considered to have closed surface shells as all regions except for the vertices form a two-dimensional network of particles in {111} facets that enclose the entire cluster.

**Anti-Mackay Clusters:** These icosahedral clusters contain more particles than Mackay clusters and display a more complex surface structure (Figure 1b).[14,16,36] Their surface shell consists of hexagonal {111} facets (green) sharing edges with rectangular {100} facets (yellow)[11,13,41] and remains closed except near the five-fold vertices (Figure 1b inset and Figure S3). The interior of an anti-Mackay cluster consists of a central Mackay core surrounded by anti-Mackay shells (Figure S1b, bottom left).[14,15] The addition of anti-Mackay shells onto the Mackay core increases packing efficiency under the constraints of spherical curvature, making them the preferred cluster geometry for medium system sizes.[14,15,17,36] The number of anti-Mackay shells can be determined by visual inspection from the geometry of the rectangular {100} facets.[14] In the example of Figure 1b, a surface rectangle five spheres wide indicates the presence of four anti-Mackay shells (light blue) twinned over the Mackay core (dark blue). Anti-Mackay clusters also have closed surface shells, as the surface consists of particles in {111} and {110} facets forming an interconnected two-dimensional network, with particles at the boundary being shared by both facets. Defects particles can reside in the five-fold vertices regions, which can be less ordered. However, excess particles are not found as adatom defects over {111} or {110} facets in experiments or simulations, as this configuration creates large gap between colloidal cluster and the droplet spherical confinement, inducing large energy penalty.

**Football Clusters:** These clusters are a distinct type for system sizes between anti-Mackay clusters and fcc clusters (Figure 1b, 50,000 colloidal particles). They have icosahedral symmetry, like (anti-)Mackay clusters, but their surface shell is not closed, like fcc clusters. Multiple discrete {111} terraces are visible as smaller hexagons stacked over larger hexagons in Figure 1c.[37] Furthermore, the {111} facets are separated by corrugated {110} facets (Figure 1c, inset). Spherical truncation of a large Mackay cluster model accurately reproduces the observed surface structure including the terraces and corrugated facets (Figure 1c model and Figure S4). The combination of twelve pentagonal vertex regions and twenty hexagonal {111} regions at the surface with the geometry of a truncated icosahedron resembles a typical stitched football, hence the term football clusters.

**Fcc Clusters:** These single-domain fcc clusters are common in large clusters beyond 100,000 colloidal particles, where the influence of finite size and confinement diminishes.[16,42–45] Figure 1d shows a fcc cluster viewed along a three-fold axis. The cluster structure matches the spherical truncation of a bulk fcc crystal (Figure 1d, model, red).[40,43,45] The surface exhibits eight hexagonal {111} facets and six square {100} facets. The surface shell of fcc clusters has terraces and steps (Figure 1d, inset) and therefore is not closed. Stacking faults in fcc clusters cost low free energy and are common.[43,45,46] For simplicity, we use the term fcc cluster for all single-domain clusters.

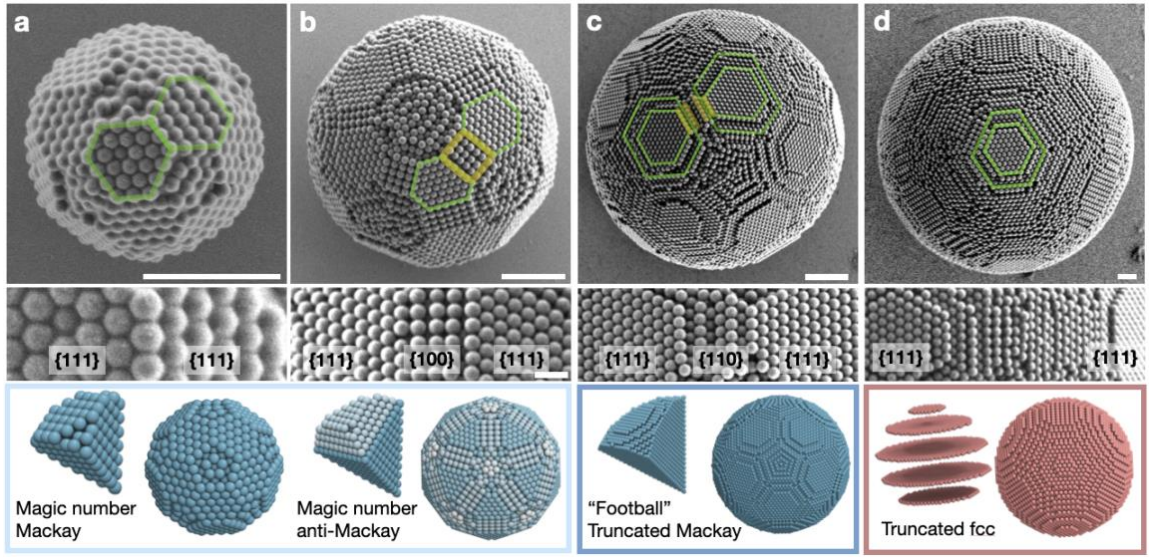

*Figure 1. Four types of colloidal clusters observed via electron microscopy for increasing system size. (a) Mackay clusters are small-sized, spherically truncated Mackay icosahedra, exposing a closed surface shell with hexagonal {111} facets (green) around five-fold axes. (b) Anti-Mackay clusters are medium-sized, spherically truncated Mackay cores with anti-Mackay shells, exposing a closed surface shell with hexagonal {111} facets (green) sharing edges with rectangular {100} facets (yellow). (c) Football clusters are large-sized, spherically truncated Mackay icosahedra, exposing a non-closed surface shell, in which hexagonal {111} terraces (green) are separated by corrugated {110} facets (yellow). (d) Fcc clusters are even larger-sized, spherically truncated fcc crystals, exposing a non-closed surface shell, in which eight hexagonal {111} facets (green) and six square {100} facets are separated by terraces and steps. Insets below show magnified views of the closed surfaced shells (a,b), non-closed surface shells (c,d) and the ideal sphere packing models together with one of the twenty constituent tetrahedral grains with Mackay structure (a), additional anti-Mackay surface termination (light blue in b), surface terraces (dark blue in c), and stacked layers of fcc the lattice (red in d). Scale bars: 2 μm.*

**Occurrence of Football Clusters and Structural Characterization**

We systematically vary the number of constituent colloidal particles from 2,000 to 400,000 particles using droplet-based microfluidics and double emulsions for enhanced control of drying conditions.[35] We analyze morphology of a large number of clusters from SEM image to statistically evaluate the occurrence of each type of colloidal cluster as a function of system size.[43] For each cluster sizes, more than 70 colloidal clusters were examined and attributed to a cluster type based on its characteristic surface features into (anti-) Mackay, football and fcc clusters (for sizes smaller than 10,000, only 30 colloidal clusters were examined as they are uniformly icosahedral). Error bars associated with the distribution of particle numbers for each cluster size are too small to be seen in the logarithmic axis of the figure (Figure S5). We exclude occasionally occurring decahedral clusters[18,46] in this analysis to simplify the analysis. As shown in Figure 2a, magic number Mackay and anti-Mackay clusters dominate for system sizes up to about 20,000 particles, after which their occurrence steadily decreases. At this size, football clusters begin to appear with increasing frequency. They reach a maximum occurrence of 60% at around 60,000 particles and disappear at about 150,000 particles. The dominance of football clusters is evidenced by lower magnification SEM images that compare clusters within the (anti-)Mackay region (Figure 2b, around 2,000 particles) and within the football cluster region (Figure 2c, around 50,000 particles). Fcc clusters are exclusively observed for systems larger than 200,000 particles, suggesting that confinement effects are negligible and bulk behavior is recovered. Notably, the football clusters form the majority species at intermediate numbers and clearly separate the regions in the number-structure phase space where (anti-) Mackay and bulk fcc structures dominate.

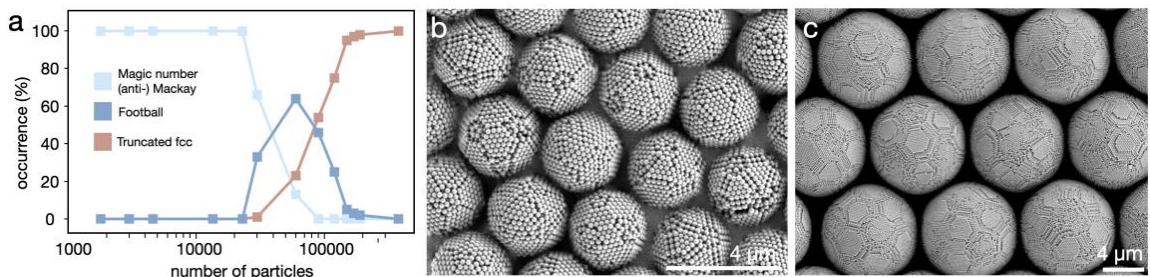

*Figure 2. Occurrence of football clusters as a function of system size. (**a**) Frequency analysis of the occurrence of magic number (anti-)Mackay clusters, football clusters, and truncated fcc clusters. At small particle numbers, (anti-)Mackay clusters are observed exclusively. At large particle numbers, fcc clusters form. Football clusters dominate at intermediate system sizes. Low magnification SEM images showing monomorphic populations of (anti-)Mackay clusters with closed surface shells (**b**) and football clusters with non-closed surface shells (**c**) as the dominating species within their respective size ranges.*

We perform X-ray nano computed tomography (nanoCT) on a selected cluster (Figure S6) to resolve the structure and internal arrangement of a football cluster (Figure 3, Supplementary Video 1). For the tomography, the cluster is glued to the top surface of a micropillar fabricated by focused-ion beam cutting, and a series of X-ray transmission images are captured during the rotation of the pillar. The three-dimensional rendering of the reconstructed cluster, shown in Figure 3a, reveals the non-closed surface shell, corroborating the impression seen in the SEM images (Figure S4). A cross-section through the reconstructed cluster along the two-fold symmetry axis reveals two opposing tetrahedral grains that converge at the cluster center (Figure 3b, marked in blue). A cross-section along the five-fold symmetry axis reveals concentric pentagonal patterns extending from the cluster surface to the center, formed by five twinned grains (Figure 3c). The crystalline domains and grain arrangements demonstrate a global icosahedral symmetry, akin to the smaller (anti-)Mackay clusters previously studied.[14–16,36,47] A magnified view along the icosahedral edges (Figure 3d) reveals a cross-sectional view of two terraces formed by the {111} facets, which are distinctly separated at the cluster surface. This corrugated surface topography is consistent with the football cluster model (Figure S4) and contrasts with the interconnected facets in the closed-shell anti-Mackay cluster (Figure S3).

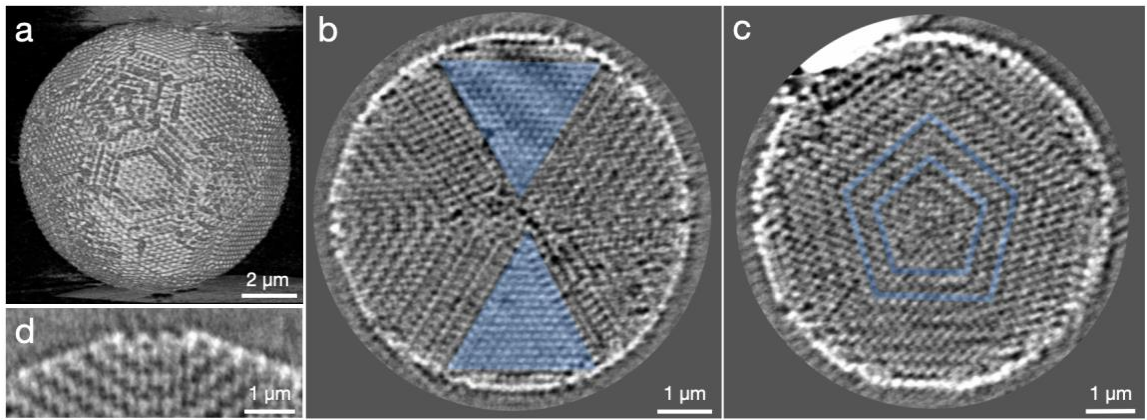

***Figure 3. Structural analysis of a football cluster by X-ray nano-computed tomography***. (***a***) *Three-dimensional reconstruction rendering of the selected cluster.* (***b***) *Cross-sectional view of the reconstructed cluster along the two-fold symmetry axis, revealing two opposing tetrahedral grains, exposing {111} facets (blue triangles to guide the eye).* (***c***) *Cross-sectional view along the five-fold symmetry axis, showing concentric pentagonal patterns formed by five twinned tetrahedral grains (blue lines to guide the eye).* (***d***) *Cross-sectional view along the icosahedral edge of the cluster, revealing corrugated surface structures between two adjacent terraces.*

**Geometric Analysis of Magic Number Clusters in Spherical Confinement**

The observation of football clusters suggests a size limit for the formation of closed-shell clusters. Given that football clusters exhibit icosahedral symmetry, distinct from the cubic symmetry of fcc clusters, this size limit cannot be simply attributed to the recovery of bulk structure. Instead, we propose that the limit arises from geometric constraints. Previous research has validated that spherical truncation of anti-Mackay sphere packing models accurately predicts the structure of magic number clusters.[14] The truncation removes all spheres beyond a specific distance from the cluster center. Increasing the truncation radius results in larger clusters with more spheres. We now explain this size limit using the sphere packing model.

Figure 4a illustrates six model substructures obtained from the sphere packing model by gradually increasing the truncation radius (Figures S1 and S3). A specific truncation radius produces the magic number cluster shown in Figure 4a(i). As the truncation radius increases (Figure 4a(ii)), more space is created between the cluster surface and the confinement, allowing extra adatom-like spheres (green) to be retained at the cluster surface. Further increase in the truncation radius forms islands of seven spheres on the {111} facets (green) and islands of two spheres on the {100} facets (yellow, Figure 4a(iii)). Such islands and terraces are not observed in experiments and simulations because they decrease entropy.[14–16,36,48] The islands grow (Figure 4a(iv)) until they merge into a new closed shell (Figure 4a(v)), in which the width of the {100} rectangular facets increases from three to four and their length decreases from six to five (Figure 4a(i) vs. 4a(v)). The anti-Mackay cluster in Figure 4a(v) is the next in a sequence of magic number clusters. A new cycle begins as the truncation radius increases further (Figure 4a(vi)).

To propose a geometric argument for a size limit for closed-shell clusters, consider two tetrahedral grains of an anti-Mackay cluster denoted ABCO' and ABCO (partially shown in Figure 4b, with details in Figure S7).[12] Spherical truncation (light red circle) produces the triangle DEF (green) and the rectangle EDGH (yellow), representing a hexagonal {111} facet and a rectangular {100} facet at the cluster surface, respectively. Two parameters determine whether additional adatom-like spheres are created by the truncation operator on these facets: the distances $D_{\{111\}} = |\text{KJ}|$ and $D_{\{100\}} = |\text{NM}|$ between the {111} and {100} facets and the curved confinement, respectively. A closed shell forms if and only if all gaps between the cluster surface and the spherical confinement interface are sufficiently small. As shown in the Supplementary Information (Figures S8-S13), the conditions $D_{\{111\}} < d_{\{111\}}$ and $D_{\{100\}} < d_{\{100\}}$ must both be met, where $d_{\{111\}}$ and $d_{\{100\}}$ are the separation distances of the {111} and {100} lattice planes. With increasing cluster size, it becomes increasingly difficult to fulfill both conditions simultaneously until, eventually, closed surface shells and thus magic number clusters are not geometrically viable anymore.

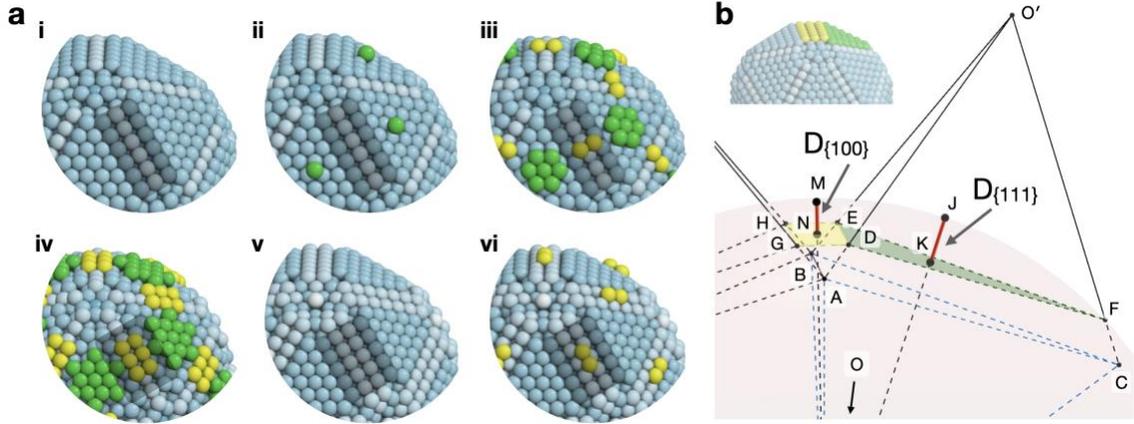

*Figure 4. Geometric construction of clusters by spherical truncation of an anti-Mackay sphere packing model. (a) Increasing the truncation radius is a strategy to construct all possible magic number closed-shell clusters. (i), A closed-shell magic number cluster with {111} and {100} facets. (ii-iv), Gradually increasing the truncation radius creates islands on the cluster surface due to the larger gap between the cluster surface and the curved confinement (green and yellow spheres) until (v), The next closed-shell magic number cluster is produced. (vi), Further increase in the truncation radius repeats the cycle. (b) Geometry of two tetrahedral grains of a Mackay core, labelled ABCO and ABCO'. The distance between the cluster surface and the confinement over the {111} and {100} facets correspond to the line segments $D_{\{111\}} = |KJ|$ and $D_{\{100\}} = |MN|$, respectively.*

**Breakdown of Magic Numbers in Colloidal Clusters**

The construction of colloidal clusters by spherical truncation of anti-Mackay sphere packing models requires specifying two parameters: the number of shells in the Mackay core (a discrete parameter) and the truncation radius (a continuous parameter). We systematically vary these parameters to construct all possible clusters and screen for those with closed surface shells, which are the magic number clusters. The size and number of anti-Mackay shells for each magic number cluster are marked as orange dots in Figure 5a. As the radius increases, more anti-Mackay shells are required to maintain closed-shell configurations, which comply more efficiently with the spherical curvature.[14,16,36] We enumerate all possible icosahedral configurations as cluster sizes increase (Figure S10-S13), and find numerically that beyond a critical radius $r/\sigma < 22$ (dashed line, $\sigma$ is the diameter of a constituent colloidal particle), no clusters can simultaneously fulfil the twin conditions that $D_{\{111\}} < d_{\{111\}}$ and $D_{\{100\}} < d_{\{100\}}$. This suggests that geometry does not permit closed-shell clusters beyond this critical cluster size.

We experimentally analyzed 140 icosahedral colloidal clusters with $3 \leq r/\sigma \leq 30$ (blue dots in Figures 5a, exemplary clusters in Figure S14). The number of anti-Mackay shells was determined visually by counting the width of the rectangular {100} facets.[14,15,43] Our experimental observations align with the geometric predictions (orange dots in Figure 5a, Figure S13). The number of anti-Mackay shells increases linearly from 0 to 7 as $r/\sigma$ increases from 3 to 23. Beyond this radius, no closed-shell colloidal clusters are observed, consistent with the prediction of the geometric model. The data confirms football clusters without anti-Mackay shells as a new type of colloidal clusters for $r/\sigma > 15$ (dotted line).

For small clusters, $r/\sigma < 5$, the gap between the cluster surface and the curved confinement is too small to fit extra particles, resulting in spherically truncated Mackay clusters with only 20 grains (Figure 1a). As the cluster radius increases, the gaps between surface facets and droplet confinement (both $D_{\{111\}}$ and $D_{\{100\}}$) increase (Figure S10). Surface reconstruction with anti-Mackay shells introduces extra twinning planes but avoids isolated islands on the cluster surface, preventing free energy penalties. The more faceted anti-Mackay geometry increases compliance with the spherical interface (Figure 1b). As the cluster radius increases further, the number of anti-Mackay shells can no longer avoid islands or terraces on the cluster surface, making anti-Mackay clusters less entropically advantageous. The preferred configuration transitions to football clusters, which are formed by truncating Mackay icosahedra and have fewer twinning planes compared to anti-Mackay clusters. The coexistence of anti-Mackay clusters and football clusters (Figure 2a, Figure 5a) in the range $15 < r/\sigma < 22$ suggests a small free energy difference between these types. Some clusters at these sizes exhibit features of both closed-shell anti-Mackay clusters and non-closed-shell football clusters mixed on the cluster surface (Figure S12, $r/\sigma = 23.2$).

We computed the free energy of hard sphere colloidal clusters in spherical confinement up to unprecedented large system sizes of $N = 20,000$ particles ($r/\sigma = 19$, Figure 5b, data up to $N = 8,000$ were taken from our previous study[14]). Simulations were performed at a packing fraction of 52% in hard spherical confinement. The fluctuation of the free energy per particle reconfirms the magic number effect.[14] For cluster radii below 15, magic number clusters with closed surface shells exist in local free energy minima. The periodically occurring free energy minima correspond to clusters with increasing numbers of anti-Mackay shells (Figure 4a(i,v)). The strength of the magic number effect, quantified by the relative depth of the free energy minima, decreases with increasing cluster radius. Beyond a cluster radius of 15, no free energy fluctuations are observed, marking the thermodynamic breakdown of magic numbers and indicating that forming closed surface shells is no longer thermodynamically favorable. This limit coincides with the emergence of football clusters (dotted line in Figure 5a and 5b).

The transition from anti-Mackay clusters via football clusters to fcc clusters implies a general trend for crystallization under an interface with decreasing curvature. The interface's influence penetrates several particles deep into the bulk.[49–52] Given that twinning costs extra free energy (volume effect) while high-coordinated surfaces reduce free energy (surface effect), hard sphere crystals under decreasing curvature may favor less twinning and more low-coordinated surfaces (Figure 5c). This suggests that free energy penalties of volumetric twinning planes outweigh those of corrugated surfaces as the interface flattens. The breakdown of magic numbers appears to be associated with the onset of recovery towards bulk properties: the disappearance of closed-shell magic clusters coincides with the system size when fcc clusters are first observed (Figure 2a). However, following this onset, fcc clusters are only observed as a minority population for a broad size range. Instead, football clusters with icosahedral symmetry dominate these intermediate sizes. Colloid crystallization in confinement is known to be driven by entropy maximization.[16] Yet it has been shown that kinetics plays an important role alongside thermodynamic considerations.[18] For example, at some specific system sizes, decahedral clusters exhibit lower free energy than icosahedral clusters but are rarely formed due to the bias towards icosahedral symmetry in the formation pathway imposed by spherical curvature of the interface.[18] Such observations at small system size suggests that at larger system sizes, kinetic effects may also contribute to the structure formation, causing a delay in bulk fcc symmetry recovery in confinement. A detailed study comparing the cluster formation pathway is necessary to explore such kinetic effects and rationalize the prevalence of icosahedral over fcc symmetry beyond size regions where fcc is thermodynamically preferred.

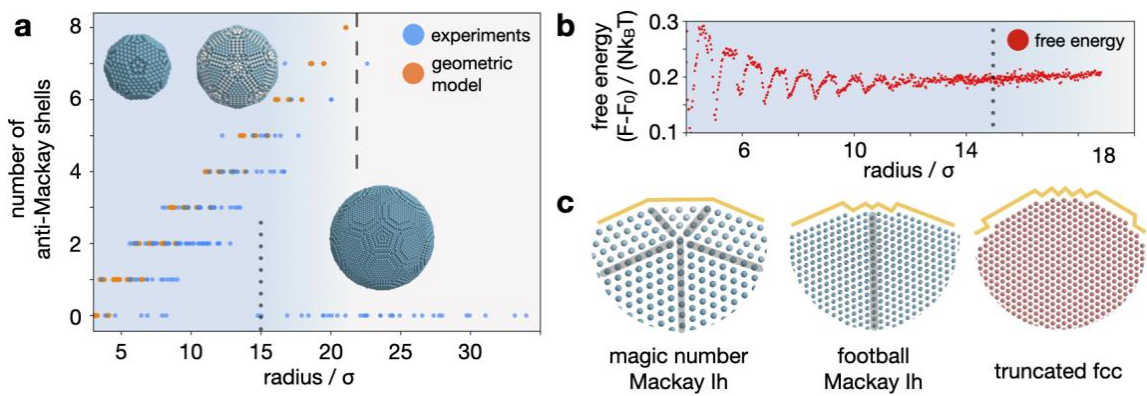

*Figure 5. Evolution of colloidal cluster geometry as a function of system size.* (**a**) Prevalence of clusters constructed from our geometric model compared to experimental observations. The number of anti-Mackay shells increases linearly through the coexistence of football clusters without anti-Mackay shells, until closed surface shell structures become geometrically impeded (dashed line). (**b**) The free energy of colloidal clusters as a function of cluster radius fluctuates periodically. The disappearance of free energy minima correlates with the appearance of football clusters in experiments (dotted line). (**c**) Colloidal clusters under a curved interface with a reducing radius of curvature and weakening confining effect form fewer grains to conform to the curvature, thereby exposing more high-index surfaces. Spheres are reduced in size for clarity.

## Conclusion

Colloidal clusters transition through four characteristic types with increasing sizes, from small, closed-shell icosahedra to bulk fcc clusters. Football clusters emerge as a distinct type at larger sizes, sharing the icosahedral symmetry of smaller magic number clusters and the non-closed surface shells of larger fcc clusters. Our investigation into football colloidal cluster reveals the breakdown of closed-shell clusters in spherical confinement.

Similar to atomic clusters, the surface energy of colloidal clusters competes with bulk energy, despite the interactions being purely entropic. While entropy maximization favors crystallization in the interior, rigid spherical confinement favors surface structures that minimize the gap between the cluster and the curved interface. These competing effects create closed-shell clusters with highly coordinated surface facets at certain sizes. Twinning improves local packing near the interface. At larger sizes, forming closed-shell clusters becomes geometrically impossible. Colloidal clusters form truncated Mackay icosahedra with fewer twinning planes, low-coordinated surfaces, and terraces. The experimental occurrence of football clusters marks this breakdown of closed-shell clusters and coincides with the attenuation of free energy fluctuations as a function of system size. This breakdown is inherent to the geometric nature of icosahedral packing.

Importantly, at the breakdown of magic numbers, icosahedral symmetry remains dominant in the form of football clusters. These clusters exist as the main population for a wide size range, from 20,000 to 100,000 particles. Bulk properties, characterized by fcc symmetry in colloidal clusters, dominate only at even larger sizes. Our previous studies reported a kinetic bias in crystallization pathways in confinement, where the spherical interface promotes and selects icosahedral symmetry over decahedral symmetry at the early stage of structure formation.[18] It is plausible that such kinetic effects also play a significant role in the transition from finite to bulk properties, facilitating the pathways to icosahedral symmetry and delaying the emergence of fcc clusters.[14,16,18,47,53]

The geometric analysis is not limited to icosahedral clusters, but can be applied similarly to decahedral colloidal clusters.[18,46] Larger decahedral clusters exhibit isolated terraces on the {111} cluster surface, reminiscent of the structural features of football clusters (Figure S14).

Our study provides general insights into the number-structure-energy relationships of finite systems up to large sizes, particularly in confinement, relevant to self-organization processes for various building blocks across different length scales.[55–68]

## Materials and Methods

**Synthesis of Polystyrene Particles**: Styrene, acrylic acid, and ammonium peroxodisulfate were purchased from Sigma Aldrich. All chemicals were used as received except for styrene, which was purified by extraction with basic water (0.1N NaOH) and filtered over alumina chromatography powder. Polystyrene (PS) particles were synthesized using ammonium peroxodisulfate as an initiator and acrylic acid as a comonomer in surfactant free emulsion polymerization. After synthesis, PS particles were kept in dialysis tubes in deionized water to remove residues from the synthesis. The deionized water was regularly renewed for a duration of one month. PS particles are electrostatically stabilized in water due to negatively charged surface functional groups.

**Production of Microfluidic Chips**: Microfluidic masters with desired patterns were produced by soft lithography. A silicon wafer was spin-coated with negative photoresist SU-8 and patterned by UV light through a photomask to create microchannels of 50 μm width. After hardening, a hexamethyldisilane anti-sticking layer was applied. Polydimethylsiloxane (Sylgard 184, Dow Corning) was mixed with a curing agent at a 10:1 ratio and poured onto the silicon master wafer. The wafer was laced in a desiccator and vacuumed to remove gas bubbles. The PDMS was cured in an 80°C oven overnight and carefully peeled off after curing. The PDMS chip was punctured with a 1 mm diameter biopsy punch (Kai Group) at inlets and outlets. The chip surface was cleaned using scotch tape and isopropanol. A clean glass slide and the cleaned chip were treated with oxygen plasma for 18 seconds at 30 W power. After plasma treatment, the PDMS chip was bonded with the glass slide. Following a one-hour rest period, the PDMS channels were flushed with Aquapel (PPG Industries) to render the inner walls hydrophobic, preventing water droplet wetting in the channel. After 30 minutes, the liquids were flushed out using compressed air.

**Fabrication of Colloidal Clusters**: A dispersion of PS particles (1 wt%, diameter of 190 nm, 244 nm, and 450 nm were used in different batches of samples) was loaded into a 1 mL syringe and pumped (Cronus Sigma syringe pump) into the PDMS chip inlet through PE tubes (0.38 mm/1.09 mm) as the dispersed phase. Perfluorinated oil (Novec Engineering Fluid HFE 7500) was used as the continuous phase and pumped into the other inlet of the PDMS chip. Krytox FSH surfactant was purchased from Costenoble Germany and used as a 0.5 wt% solution in HFE 7500 oil. The flow rate of the water phase was controlled between 100 and 1000 μL/hour, while the oil phase was controlled at 100 μL/hour to generate monodisperse droplets of different sizes. Droplets were collected by a pipette tip at the outlet of the microfluidic chip and transferred to a clear glass vial (1.5 mL) for storage. For polydisperse droplet generation, 10 μL of PS particle dispersion was pipetted into a clear glass vial (1.5 mL) containing HFE 7500 oil with 0.1% PFPE-PEG-PFPE surfactant. After sealing the vial with a cap, it was shaken vigorously by hand. All droplets, including those produced by microfluidic chips, were stored at room temperature with parafilm covering the vial opening. Both the water and oil phases evaporate, but a higher oil phase volume in the glass vial ensures that consolidated colloidal clusters after droplet drying were readily dispersed in the oil. To ensure uniform droplet drying process, colloidal clusters were also fabricated by consolidating colloid dispersion in double emulsion droplets. Single water-in-oil emulsions produced by PDMS microfluidics described in previous step were vortexed in 50 mL centrifuge tube in 5 wt% SDS aqueous solution for 30 seconds at 2500 rpm to create uniform water-in-oil-in-water double emulsion. To ensure clusters reach equilibrium in droplets, we perform slow evaporations of droplets. For single emulsion droplets containing the polystyrene particle dispersion, droplets were kept in a 1mL glass vial sealed with parafilm, three holes were made in the parafilm sealing with a 4 mm needle to allow evaporation of water. Typically, 100 μL of water phase is dispersed in 400 μL of oil. The glass vial is placed without agitation for approximately a week at room temperature to ensure slow evaporation, previously determined to be approximately 0.2 mg/hour.[14] For double emulsions, the inner water core is shrunken within 7 days by the osmotic pressure difference due to the presence of free sodium dodecyl sulfate (SDS) in the continuous water phase (5 wt.% SDS were used[35]). We compare colloidal clusters fabricated via both single and double emulsions with simulated structures and model construction to ensure the resulting clusters have reached equilibrium and crystalline structures. Clusters presented in Figure 2a were fabricated via double emulsion method, in Figure 5a via single emulsion method.

**Characterization of Colloidal Clusters**: For characterization, drops of HFE 7500 oil containing the colloidal clusters were drop-cast onto a clean silicon wafer. After ten minutes, the HFE 7500 oil evaporated from the wafer, depositing colloidal clusters to be examined by scanning electron microscope (ZEISS Gemini Ultra 50). The number of particles of clusters is obtained by measuring the cluster sizes in SEM and assuming particle density of 0.69 for icosahedral clusters and 0.74 for fcc clusters. A ZEISS Xradia 810 Ultra X-ray microscope equipped with a 5.4 keV rotating anode Cr-source and a Zernike phase ring for positive phase contrast imaging allowed imaging of 16 × 16 µm large areas with optical resolutions down to 50 nm (pixel size of 16 nm). Several clusters of one ensemble were prepared on a silicon surface and imaged by SEM to find the football cluster sample. The chosen sample (Figure S6) was transferred onto a needle tip in the pre-alignment light microscope of the X-ray microscope for 360° tomography without missing wedge. A tilt series with a total number of 901 transmission images was recorded with an acquisition time of 400 s/frame. The image series was aligned along the rotational axis by metrology tracking of temperature dependent shift, and the complete two-dimensional dataset was relocated to fit the reconstruction geometry. Acquisition and alignment were performed in the native ZEISS microscope software (XMController and Scout&Scan), and three-dimensional reconstruction was performed employing the Simultaneous Iterative Reconstruction Technique (SIRT) algorithm implemented via an in-house Python script based on the Astra-Toolbox.[68–70] Arivis Vision4D and InViewR were used for visualization in 2D and by virtual reality. Geometric calculations were performed using a Python code. Details of geometric model are described in *SI Materials and Methods*.

**Simulation of Colloidal Clusters and Free Energy Calculation**:

Our previously developed two-step simulation procedure was utilized to replicate colloidal clusters observed in the experiments. In the first step, colloidal particles were modelled as hard spheres with diameter $\sigma$ in a hard spherical confinement of radius $R$.[14] Evaporation was simulated with event-driven molecular dynamics in the isochoric (NVT) ensemble by gradually increasing the packing fraction from $\phi = 0.48$ to $\phi = 0.55$ in steps of 0.001. At each step, a simulation was performed for a duration of $\triangle t* = 500$, which was sufficient for a robust and reliable icosahedral ordering. In the second step, to mimic the capillary forces that consolidate colloidal clusters during the final stage of droplet drying, the interaction of the hard sphere was replaced with the Morse potential $V(r) = D_0(e^{-2\alpha(r-r_0)} - 2e^{-\alpha(r-r_0)})$, wherein $r_0 = \sigma$, $D_0 = 1$, $\alpha = 10$, and energy was minimized with the fast inertia relaxation engine (FIRE) implemented in HOOMD-blue.[71,72] A force tolerance per particle of $10^2$ and an energy tolerance of $7 \times 10^7$ was used.

We extended our previously used free energy method without modification to larger clusters.[14,15] The calculation of free energy is based on the Frenkel-Ladd method[73] with inclusion of particle swap moves[74] in a Monte Carlo simulation. Einstein crystal was taken as a reference system with harmonic springs, $V(r) = \lambda(r - r_0)^2$, where $\lambda$ is the spring constant and $(r - r_0)$ is the measure of displacement from the spring anchor points $r_0$. The spring constant was increased logarithmically in discrete steps over the range from $10^{-5}$ to $10^5$. Absolute free energies were obtained after subtracting the bulk contribution $F_0(N, \phi)$ to the free energy.

## Data Availability Statement

The data (including raw and metadata) underlying this study are openly available in Zenodo at zenodo.org/records/14616649.

## Acknowledgments

This work was supported by Deutsche Forschungsgemeinschaft (DFG) through the projects EN 905/2-1 and VO 1824/7-1 to M.E. and N.V., respectively. Support by the Deutsche

Forschungsgemeinschaft under project ID 416229255 – SFB 1411 is also acknowledged. Scientific support and HPC resources provided by the Erlangen National High Performance Computing Center (NHR@FAU) under the NHR project b168dc are gratefully acknowledged. NHR funding is provided by federal and Bavarian state authorities. NHR@FAU hardware is partially funded by DFG – 440719683.

*Supplementary Information for*
# Breakdown of Magic Numbers in Spherical Confinement

Junwei Wang[1,4]*, Jonathan Martín-González[1], Lukas Römling[1], Silvan Englisch[3], Chrameh Fru Mbah[2], Praveen Bommineni[2], Erdmann Spiecker[3], Michael Engel[2]*, Nicolas Vogel[1]*

[1]*Institute of Particle Technology, Friedrich-Alexander-Universität Erlangen-Nürnberg, 91058 Erlangen, Germany*
[2]*Institute for Multiscale Simulation, Friedrich-Alexander-Universität Erlangen-Nürnberg, 91058 Erlangen, Germany*
[3]*Institute of Micro- and Nanostructure Research and Center for Nanoanalysis and Electron Microscopy, Friedrich-Alexander-Universität Erlangen-Nürnberg, 91058, Erlangen, Germany*
[4]*Max Planck Institute of Colloids and Interfaces, 14476, Potsdam, Germany*
***Emails:** junwei.wang@mpikg.mpg.de, michael.engel@fau.de, nicolas.vogel@fau.de

# 1. Geometry of Icosahedral Colloidal Clusters

## 1.1. Sphere Packing Model of Extended Mackay Icosahedra

The sphere packing model uses identical spheres, which are marked with different colors in the model to facilitate identification of the different structural components. The model is based on a Mackay icosahedron, consisting of 20 tetrahedral grains (Figure S1a). To construct colloidal clusters with surface anti-Mackay shells, twinned tetrahedral grains are added over the faces, edges, and vertices of the Mackay Icosahedron (Figure S1b-d). A detailed explanation of the model can be found elsewhere[1]. The key to accurately reproduce colloidal clusters observed in the experiment is the spherical truncation, which removes all spheres outside of a given truncation radius and therefore approximates the role of the spherical confinement (Figure S1e). The model icosahedral cluster can be described by two numbers, $m$, the number of interior Mackay shells, and $a$, the number of surface anti-Mackay shells. The surface patterns of icosahedral clusters are determined uniquely by these two numbers. Therefore, the surface of a cluster can be used to infer its interior structure. It is important to note that once $m$ is fixed, the extended Mackay icosahedron model is fixed. Using different truncation radii can then create substructures of different numbers of anti-Mackay shells $a$.

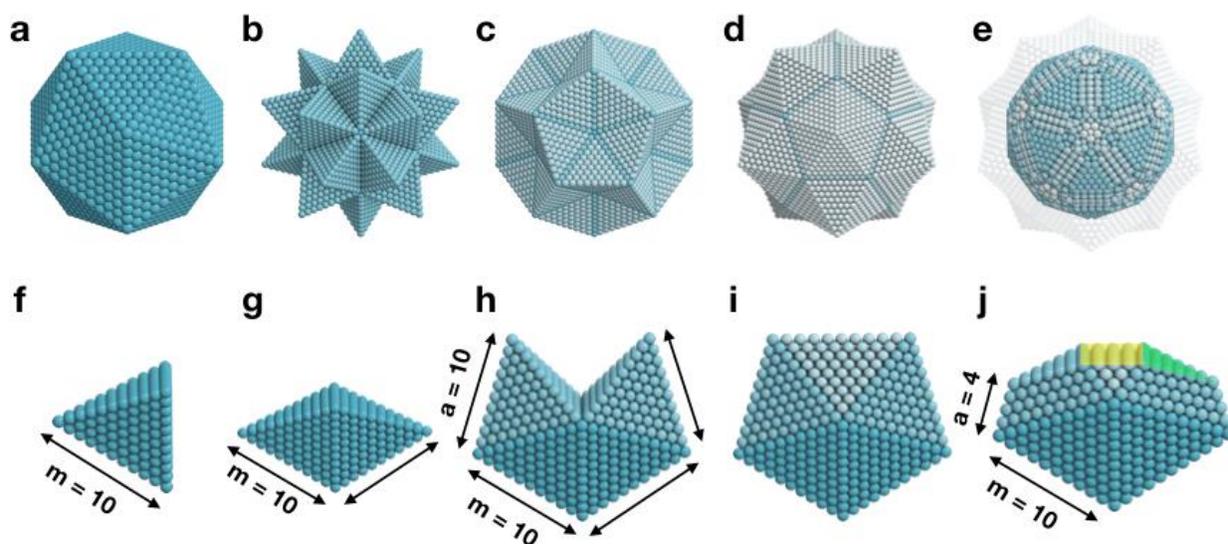

*Figure S1. Extended Mackay icosahedron model.* **a**, Sphere packing model of Mackay icosahedron, constructed by addition of concentric icosahedral shells around a central sphere in ABCABC sequence, making up 20 slightly deformed fcc tetrahedral grains. **b**, 20 extra tetrahedral grains are added to the faces of the Mackay icosahedron. **c**, 30 more tetrahedral grains are added over the edges of the Mackay icosahedron. **d**, 60 most deformed tetrahedral grains are added over the vertices of the Mackay icosahedron. The extra tetrahedra in (b,c,d) make up the surface twinning shells over the Mackay icosahedron. Once the size of the Mackay icosahedron is fixed, the surface shells are also determined. **e**, Spherical truncation mimics the effect of rigid droplet interface during droplet drying and cluster formation, which removes spheres outside the truncation radius. Spherical truncation of varying radius of different models creates model substructures, which describe small, medium, and large colloidal clusters. **f**, One piece of tetrahedral grain in (a) in the interior Mackay region with $m = 10$ layers. **g**, Two adjacent grains in the Mackay icosahedron twinned together. **h**, Two tetrahedral grains in the anti-Mackay regions added onto the two Mackay tetrahedral grains. The added grains have the same size $a = 10$ layers, corresponding to (b). **i**, Additional grain in the anti-Mackay region added between the gap, corresponding to (c). The geometric model in Figure 2 and Figure S5 are abstracted from here. **j**, Spherical truncation removes some spheres in the anti-Mackay region, corresponding to €. After truncation, only $a = 4$ anti-Mackay shells remain, the {111} (green) and {100} (yellow) facets are exposed.

## 1.2. Three Types of Icosahedral Colloidal Clusters

Depending on the system size (cluster radius), three types of icosahedral colloidal clusters are identified by their geometric features. The small clusters are truncated Mackay icosahedral clusters (Figure S2), the medium clusters are truncated extended Mackay icosahedral clusters with anti-Mackay shells (Figure S3), the large clusters are football clusters also consisting of truncated Mackay icosahedral clusters (Figure S4). However, for small clusters, the surface structure is a closed shell with interconnected, hexagonally close-packed {111} facets, while large clusters do not exhibit a closed surface shell and form terraces with both {111} and {110} facets.

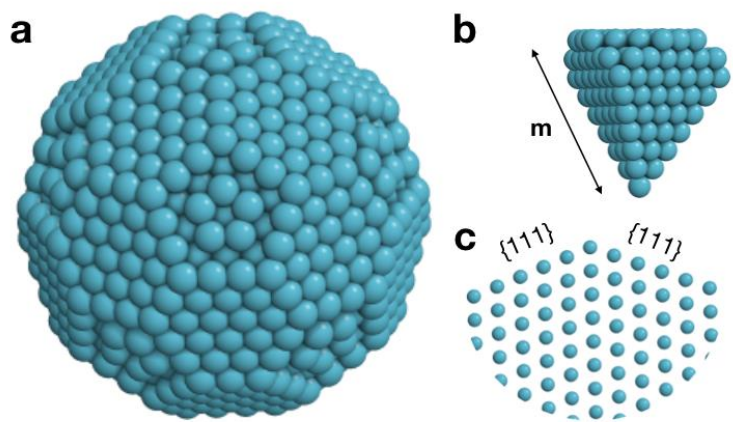

*Figure S2. Sphere packing model of small icosahedral colloidal cluster, generated by spherical truncation of small Mackay icosahedron. a, The cluster surface is tiled with hexagons at {111} facets. b, One Mackay tetrahedral grain. c, Cross-section of the cluster at the surface regions underlines the closed-shell nature of the icosahedral cluster. Spheres are reduced in size for clarity.*

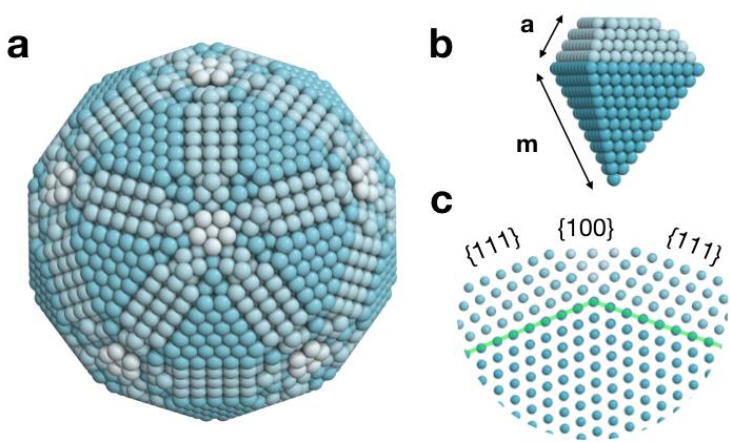

*Figure S3. Sphere packing model of medium icosahedral colloidal cluster, generated by spherical truncation of the extended Mackay icosahedron. a, The cluster surface is tiled with hexagons and rectangles at {111} and {100} facets. b, The grain consists of $m$ Mackay shells and $a$ anti-Mackay shells, in this case, $a = 4$ and $m = 10$. c, The closed-shell structure of the surface region is shown in the cross-section of the cluster on the right. Spheres are shrunk for clarity, different color represent their origin from*

*different tetrahedral grains shown in Figure S1a-c. The twinning plane separating the Mackay core from the anti-Mackay surface shells is marked by a green line. Spheres are reduced in size for clarity.*

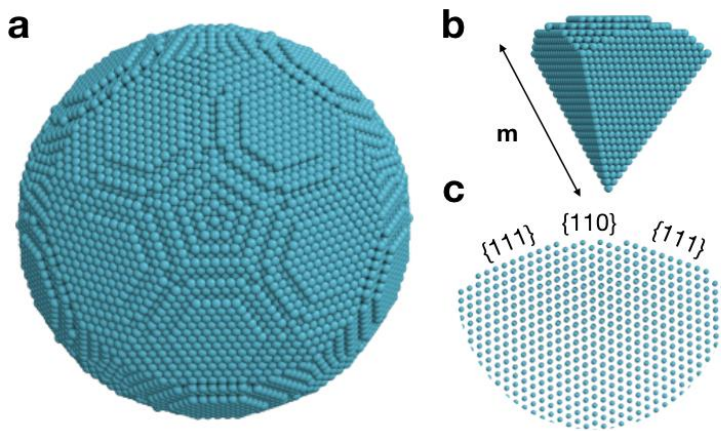

***Figure S4. Sphere packing model of large, football icosahedral colloidal cluster, generated by spherical truncation of large Mackay icosahedron. a***, *The cluster surface shows terraces at {111} facets and {110} facets.* ***b***, *The grain consists of only Mackay shells, but exhibits terraces at the surface.* ***c***, *The first few shells at the cluster surface are disconnected, as shown in the cross-section of the cluster on the right. Spheres are reduced in size for clarity.*

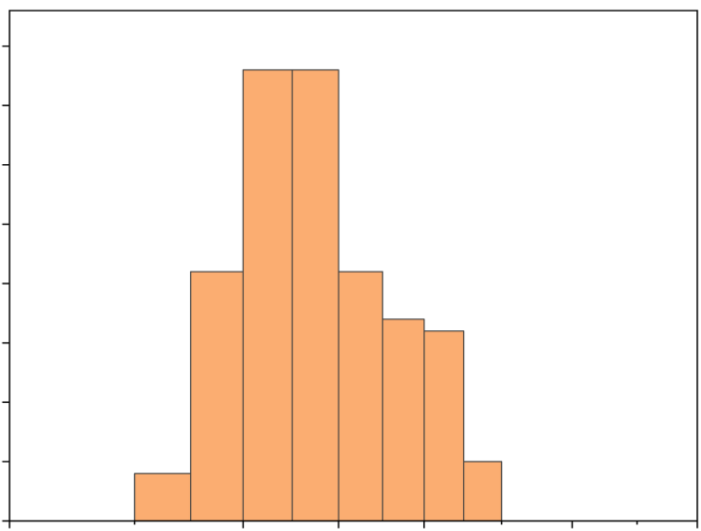

***Figure S5. Standard deviation of colloidal cluster sizes observed in SEM.*** *Standard deviation for a sample identified as 90,000 particles is 9.8% in the number of particles. Such value is too small to be visible in logarithmic axis in Figure 2. The data is collected to measure size deviation, its cluster types are not identified for use in Figure 2.*

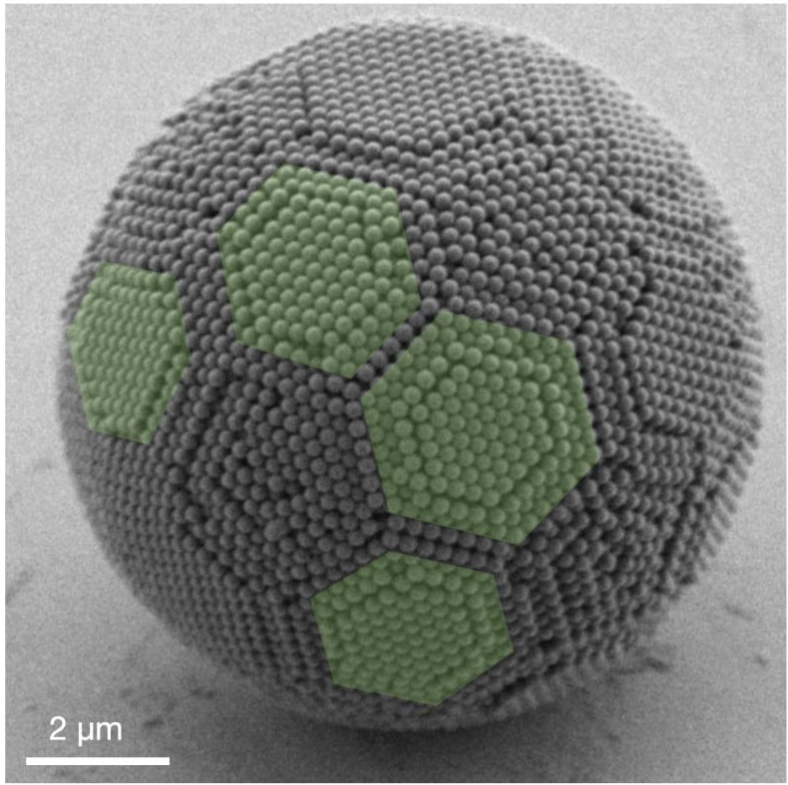

*Figure S6. Selected football cluster sample, in which the 3D Xray NanoCT tomography was performed to obtain Figure 3 and Supplementary Video 1.*

### 1.3. Geometric Model

The ideal Mackay icosahedron model is a non-crystallographic arrangement of identical spheres. The 20 tetrahedral grains making up the icosahedron are not regular but slightly deformed. This deformation is even more pronounced in additional tetrahedral grains used to construct the anti-Mackay shells.[1,2] For the theoretical prediction of the breakdown of magic number clusters, we abstract the sphere packing model into a purely geometric model (Figure S7). The first error comes from replacing the deformed tetrahedral grain with regular tetrahedra. The second error comes from replacing spheres with volume by mathematical points. Another consideration is that while the spherical droplet confinement may be slightly flexible at the end of droplet drying, our spherical truncation is rigid. However, our analysis shows that these errors only shift the predicted critical value at which shell closure fails but cannot prevent it from occurring. The failure of shell closure is inevitably rooted in the geometry of the icosahedron.

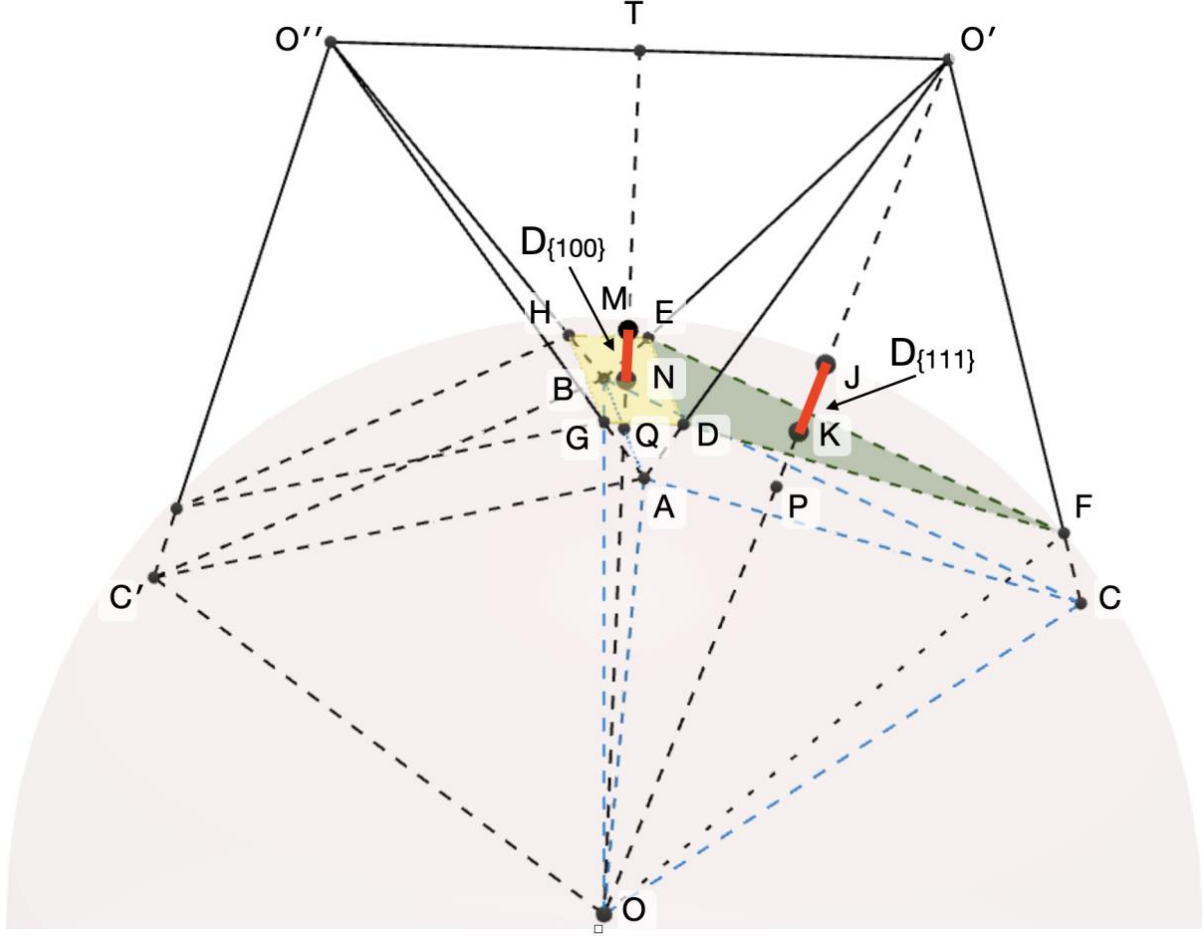

*Figure S7. Geometric model showing five tetrahedra of the cluster model and the truncation sphere (in light red). The twinned tetrahedra ABCO and ABC'O represent two adjacent tetrahedral grains in the Mackay icosahedron (Figure S1a). Tetrahedra ABCO' and ABC'O'' represent the grains over the face of Mackay icosahedron (Figure S1b). Tetrahedra ABO'O'' represent the gap over the edge (Figure S1c). The points D, E, F, G, H are intersections of the truncation sphere with radius OF and the tetrahedron. These points are at the cluster surface after truncation. Line segment JK and MN are the distance between the truncation sphere and the cluster surface, represented by $D_{\{111\}}$ and $D_{\{100\}}$.*

The regular tetrahedron $ABCO$ has an edge $OC$ of length $s$, twinned with the tetrahedron ABCO'. The spherical truncation centers at point $O$ and intersect the tetrahedron ABCO' at point $F$. The radius $OF$ has length $R$, and the line segment $FC$ has length $x$. For the triangle $FCO$, according to laws of cosines,

$$R^2 = x^2 + s^2 - 2xs \cos \angle FCO,$$

where the angle $\angle FCO$ equals $\angle O'CO$ and equals two times the tetrahedron face-vertex-edge angle $\alpha = \cos^{-1}(1/\sqrt{3})$. Hence,

$$d(FC) = x = s \cos 2\alpha \pm \sqrt{(s \cos 2\alpha)^2 + R^2 - s^2}.$$

As the truncation radius $OF$ must be larger than $OC$ and smaller than $OO'$ to have an intersection,

$$x = s \cos 2\alpha + \sqrt{(s \cos 2\alpha)^2 + R^2 - s^2},$$

where $s < R < 2s \sin \alpha$.

The line segment $OJ$ is normal to the triangle $ABC$ and $DEF$, and has the length of the truncation sphere radius. Hence, the distance between truncation sphere and cluster surface is

$$d(JK) = d(OJ) - d(OP) - d(PK),$$

where $d(OJ) = R$, $d(OP) = s\sin\alpha$, which is the height of the tetrahedron $ABCO$. $d(PK)$ is the distance between triangle $ABC$ and $DEF$, which equals to the distance from point $F$ to the triangle $ABC$. Therefore, the distance between the truncation sphere and cluster surface can be expressed by the tetrahedron and the truncation radius,

$$d(JK) = R - s\sin\alpha - x\sin\alpha,$$

$$d(JK) = R - s\sin\alpha - \left(s\cos 2\alpha + \sqrt{(s\cos 2\alpha)^2 + R^2 - s^2}\right)\sin\alpha,$$

where $s < R < 2s\sin\alpha$. Similarly, the distance between truncation sphere to cluster surface at the rectangle $DEHG$ can be expressed as:

$$d(MN) = d(OM) - d(OQ) - d(QN),$$

where the length of $OM$ equals the truncation radius $R$, the length of $OQ$ is the distance between tetrahedron vertex $O$ to midpoint $Q$ in the opposing edge $\sqrt{3}s/2$.

To simplify the calculation, we consider the tetrahedron ABO'O'' to be a regular tetrahedron, neglecting that the dihedral angle at the edge $AB$ is in fact expanded by about 7.4°. The length of line segment $QN$ is the distance between rectangle $DEHG$ and midpoint $Q$ in $AB$. Due to triangle similarity, it follows:

$$\frac{d(QN)}{d(QT)} = \frac{d(AD)}{d(AO')},$$

where the length of $QT$ is the edge-to-edge distance in the tetrahedron, $s/\sqrt{2}$, the length of $AO'$ is the edge of tetrahedron $s$, $d(AD) = d(CF) = x$. Hence,

$$d(MN) = R - \frac{\sqrt{3}}{2}s - \frac{x}{\sqrt{2}},$$

$$d(MN) = R - \frac{\sqrt{3}}{2}s - \frac{s\cos 2\alpha + \sqrt{(s\cos 2\alpha)^2 + R^2 - s^2}}{\sqrt{2}},$$

where $s < R < 2s\sin\alpha$.

For sphere packing (diameter of each individual sphere equals 1) in the face-centered cubic crystal, the distance between {111} plane is $\sqrt{6}/3$, the distance between {100} plane is $\sqrt{2}/2$. The condition to ensure no additional sphere can fit into the gap between cluster surface and truncation sphere is

$$d(JK) \leq \frac{\sqrt{6}}{3} \quad \text{and} \quad d(MN) \leq \frac{\sqrt{2}}{2}.$$

Discretizing the edge length of tetrahedra $s$ to be an integer at or above 3 (Mackay icosahedron of size 3 to infinitely large), the radius of truncation sphere is in the range $s < R < 2s\sin\alpha$. Solving the above inequality gives the phase diagram in Figure 3 and Figures S7-S9.

### 1.4. Geometric analysis for closed shell clusters

From the above derivation, we can now analytically express different important geometric quantities as a function of the radius of the cluster. We fix the size of the extended Mackay icosahedron model as $m = 10$ (Figure S1, or $OC$ in Figure S7) and explore how the geometric quantities vary as a function of truncation radius.

First, we show how the area of different facets in the cluster surface, created by the spherical truncation, changes with the truncation radius. The triangle (green line) represents the {111} facets at the face of the icosahedral cluster, the rectangle (yellow line) represents the {100} facets at the edge of the icosahedral cluster and the pentagonal region (grey line) at the vertices of the icosahedral cluster. As the truncation radius increases, the area of icosahedral faces (the {111} facets) decrease monotonically as truncation radius increases, the area of edges (the {100} facets) reaches maximum before decreasing to zero, and the area of vertices increases monotonically.

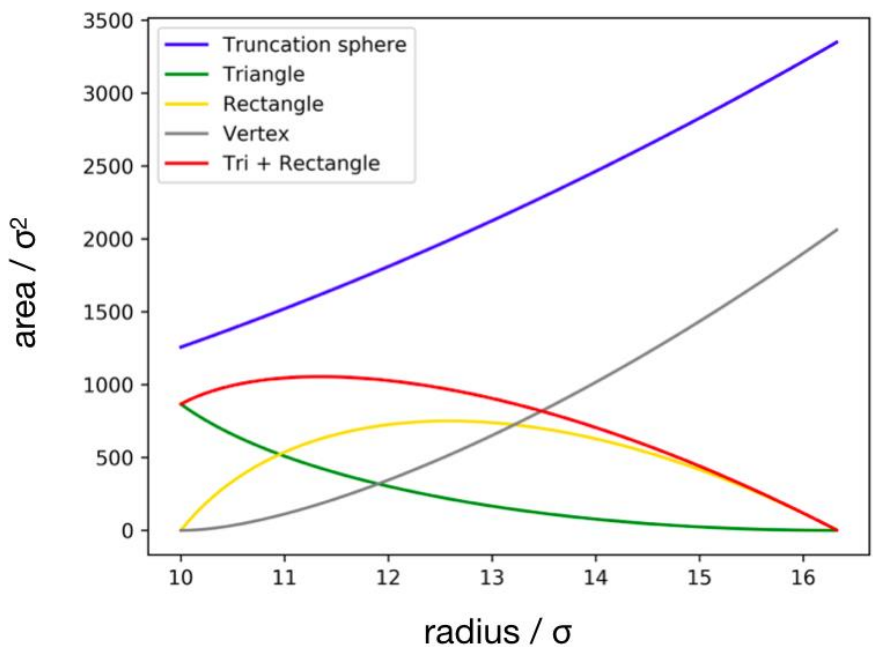

*Figure S8. Area of cluster surface as a function of spherical truncation radius.* While the area of the 20 surface triangles (green line in Figure 1 and triangle DEF in Figure S7) decreases monotonically to zero with truncation radius, the area of the 30 surface rectangles (yellow line in Figure 1 and rectangle EDGH Figure S7) reaches a maximum before it decreases to zero. The area in the vertices region in the cluster increases continuously (grey line).

Observation from experiments and simulations,[1,3,4] as well as geometric analysis of grain deformation in the icosahedral cluster[2] indicate that the vertex regions are subject to more disorder, higher strain, and lower-coordinated spheres. Therefore, a large truncation radius resulting in large area of vertices is unfavorable. We have only observed clusters, in experiments and simulations, whose surface rectangles' length (edges shared with hexagons over icosahedral faces) is no shorter than their width (edges shared with pentagon over icosahedral vertices). We call this an empirical edge rule for icosahedral clusters constructure in spherical confinement.

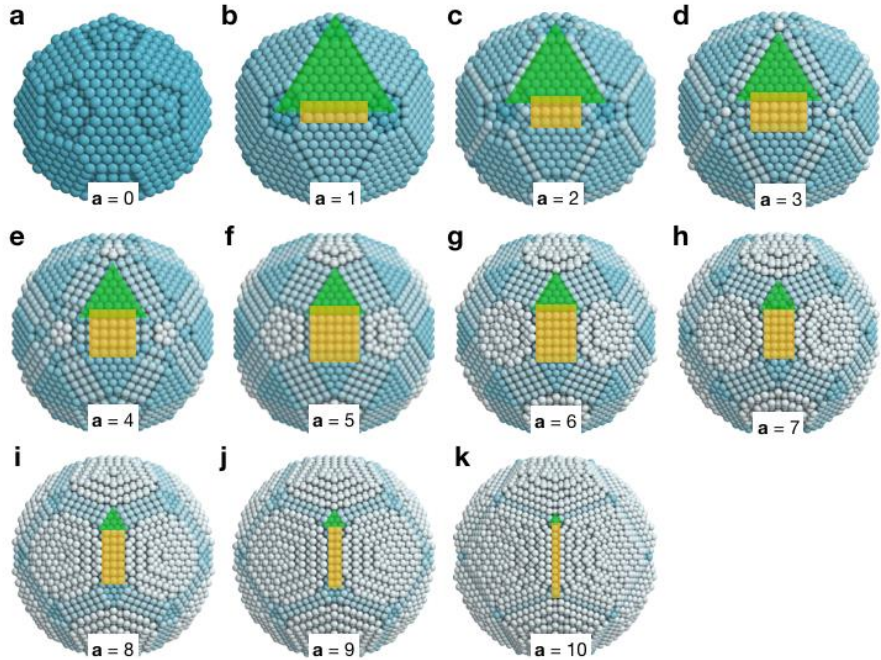

*Figure S9. Selection of closed-shell substructures from extended Mackay icosahedron ($m = 10$) generated by spherical truncation without applying the edge rule.* *This illustrates the failure to form closed shell for one exemplary icosahedral cluster ($m = 10$). The parameter $a$ indicates the number of anti-Mackay shells after the truncation. The area of icosahedral faces (green) decreases continuously, and the area of edges (yellow) has a maximum (Figure S8).* ***a-d****, These clusters are observed in experiments and simulations, in all cases the length of the rectangle (shared with hexagons) are greater than the width (shared with pentagons).* ***e-k****, These clusters are not observed. The boundary condition of $D_{\{111\}}$ and $D_{\{100\}}$ only regulate the closed surface shell structures over the faces and edges, but not over the vertices, which at large truncation radius yields large area of unfavorable vertices regions.*

We then show how the gap between the cluster surface and the curved spherical interface changes as the truncation radius increases, both over the icosahedral faces (line segment *JK* or $D_{\{111\}}$ in Figure 4, Figure S7) and edges (line segment *MN* or $D_{\{100\}}$ in Figure 4, Figure S7). As the truncation radius increases, the gap between the cluster surface and the truncation sphere over the icosahedral faces (the {111} facet) decreases continuously, but over the icosahedral edges (the {100} facet) the gap first decreases but increases slightly at large radii. This means that including more anti-Mackay shells, as a result of increasing truncation radius, is always favorable over the icosahedral faces (to reduce the gap). However, over the icosahedral edges, having too many anti-Mackay shells increase the gap to the curved spherical interface at some point. For sufficiently large models, when $m$ is sufficiently large, this opposing trend prohibits the gap over the faces $D_{\{111\}}$ and edges $D_{\{100\}}$ to be simultaneously shorter than their critical length at any truncation radius. Therefore, the failure of icosahedral shell closure is inevitable at large system sizes.

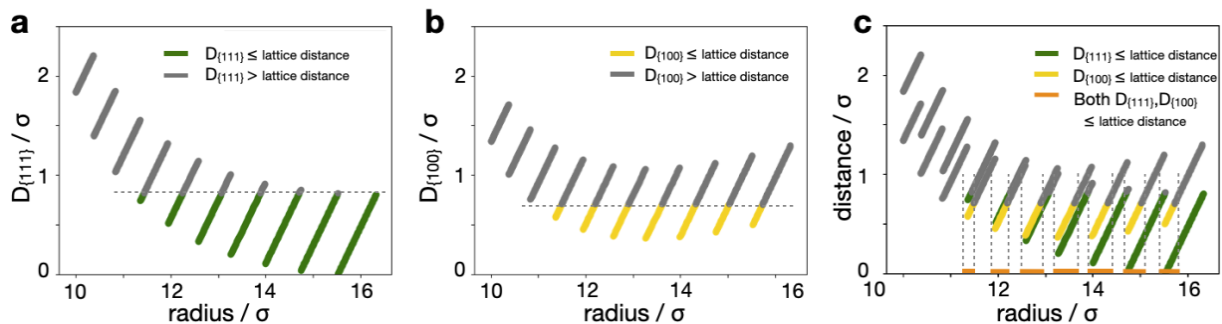

*Figure S10. Gap distance between cluster surface and spherical confinement. a, Numerical calculation of the gap between the cluster surface and the truncation sphere over the icosahedral faces (length of line segment JK or $D_{\{111\}}$ in Figure 4b and Figure S7) as a function of the truncation radius. The grey color represents truncation radii where JK or $D_{\{111\}}$ is greater than its critical distance, unable to form closed-shell clusters (where extra sphere can fit in the gap). Green represents truncation radii where JK or $D_{\{111\}}$ is shorter than the critical length, forming closed-shell clusters. The critical length is determined by the distance between {111} crystal planes in a fcc lattice. The increase of truncation radius (OF in Figure S7) is gradual, but the number of shells is a discrete integer. Hence the gap distance $D_{\{111\}}$ is not a continuous function of truncation radius. The shown calculation is based on an extended Mackay icosahedron ($m = 10$), where the spherical truncation includes at most 10 surface shells ($a = 10$) in the substructures, hence the 10 individual segments in the diagrams. b, Numerical calculation of the gap between the cluster surface and the truncation sphere over the icosahedral faces (length of line segment MN or $D_{\{100\}}$ in Figure 4b and Figure S7) as a function of the truncation radius. Note that here the critical length is the distance between {100} crystal plane in a fcc lattice. c, Closed-shell clusters can only exist when both JK and MN are simultaneously shorter than their critical length, only certain specific truncation radii satisfy this condition (marked by orange line segments at the x-axis). The orange area indicates cluster radius where closed-shell clusters are geometrically allowed.*

From Figure S10a, S10b the gaps between cluster facets and the curved truncation interface can be calculated. Only when both the gaps over the icosahedral {111} faces and {110} edges fall below their critical length, the distance between crystal planes, can the cluster have a closed surface shell. Otherwise adatom-like defect can fit in the gap between cluster surface and confinement. In Figure S10c, the range of cluster radii, which allows shell closure, is shown at the x-axis in orange. We replot the cluster radii that allows closed shell structure for further analysis in Figure S11.

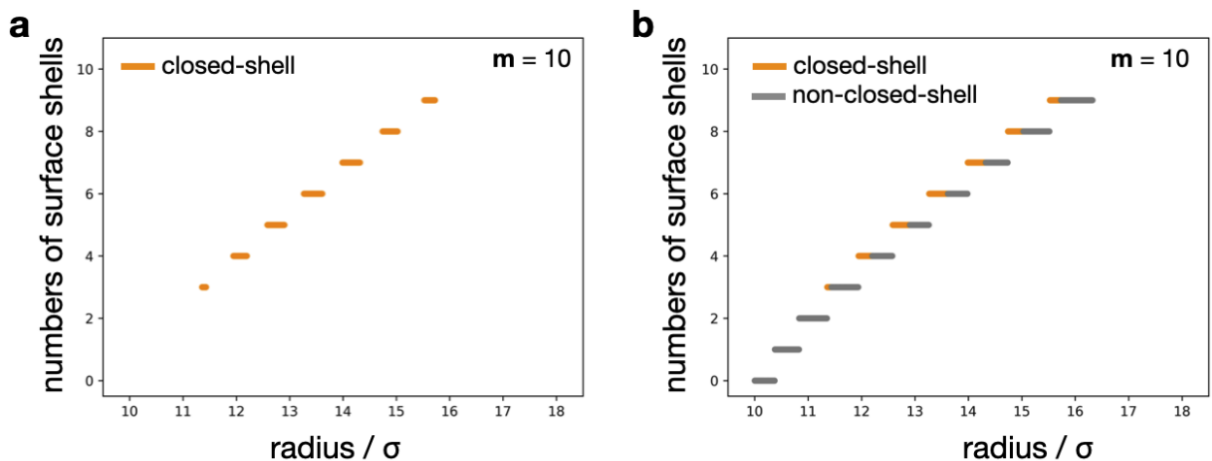

*Figure S11. The clusters with closed surface shell as a function of truncation radius based on extended Mackay icosahedron ($m = 10$). a, The cluster radii, which allow closed-shell clusters are shown in orange. The number of closed surface shells (the anti-Mackay shells) increases with cluster radii. b, Non-*

*closed-shell cluster radii are marked by grey. The range of orange and grey together makes up the entire range of cluster radii from the spherical truncation of the model ($m = 10$). Only some truncation radii can produce closed surface shells.*

The previous calculation enumerates all possibilities for closed surface shells for all cluster radii based on substructures generated by an extended Mackay icosahedron ($m = 10$). However, at a given cluster radius, there are several different possible icosahedral clusters whose interior structure consists of different number of Mackay $m$ and anti-Mackay shells $a$.[1,3] In other words, different $m$ value can be used to generate clusters of similar sizes. We now use the size of the extended Mackay icosahedron $m$ as a variable to enumerate all possible closed-shell clusters, first as an example with $m = 10$ and $m = 11$ (Figure S12), then with all integer values for $m$ (Figure S13). As $m$ can be infinitely large, the number and the size of icosahedral cluster can be infinitely large, too. However, the closed-shell clusters can only occur in a limited region of cluster sizes, regardless of the configuration, as shown by the orange areas in Figure S13a,b. The area covered by grey dots are possible icosahedral configurations (based on icosahedron core with $m$ shells); area covered by orange dots are closed shell structures. We further apply empirical edge rule (Figure S9, that in the rectangular region in the anti-Mackay cluster (Figure 1, S3), the length must not be smaller than the width). This has been the observation of all clusters found in experiments and simulations. Applying the edge rule (Figure S9) reduces the number of shells in the anti-Mackay layers (Figure S13c, S13d). The model predicts a range of geometrically possible closed-shell clusters (Figure S13c, orange) that coincides with experimental observations (Figure S13d, blue). We note that even without applying the edge rule, the geometric model predicts a breakdown of closed-shell configurations, albeit at slightly larger system sizes (Figure S13c).

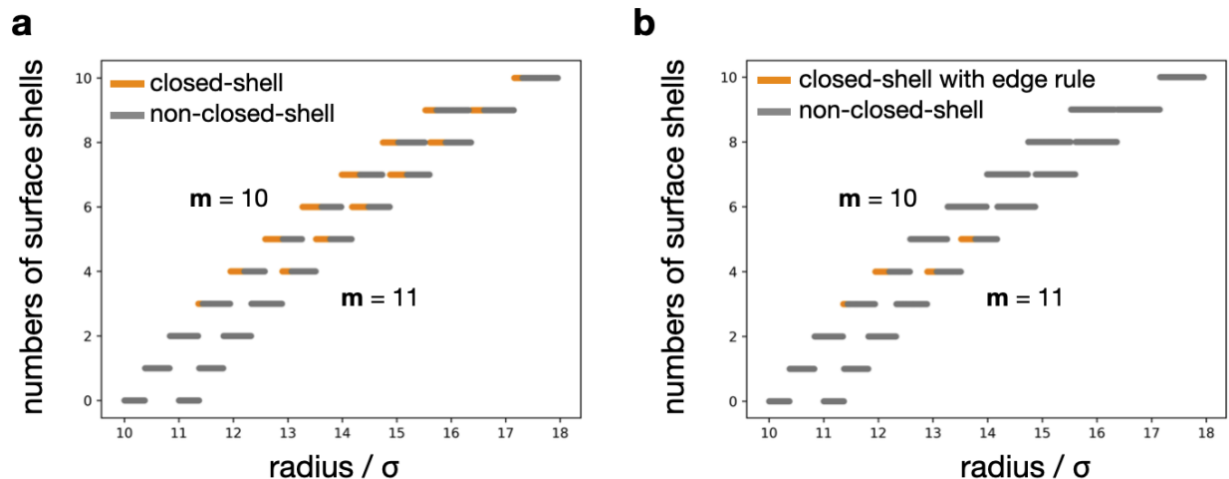

*Figure S12. The clusters with closed surface shell as a function of truncation radius based on the extended Mackay icosahedron ($m$ = 10 and 11). a, Similar to Figure S11a, the cluster radii allowing shell closure are marked in orange. Here, the results from model $m = 11$ are superimposed. Note that for any given radius there is more than one possible icosahedral cluster with different number of shells. Similarly, for any number of surface shells, there is more than one cluster radius, which allows such shell configuration. b, Applying an empirical edge rule (Figures S9), which limits the width (shared with pentagon at the vertices) to be shorter than the length of the rectangle surface tile (shared with hexagons) reduces the occurrence of closed-shell clusters.*

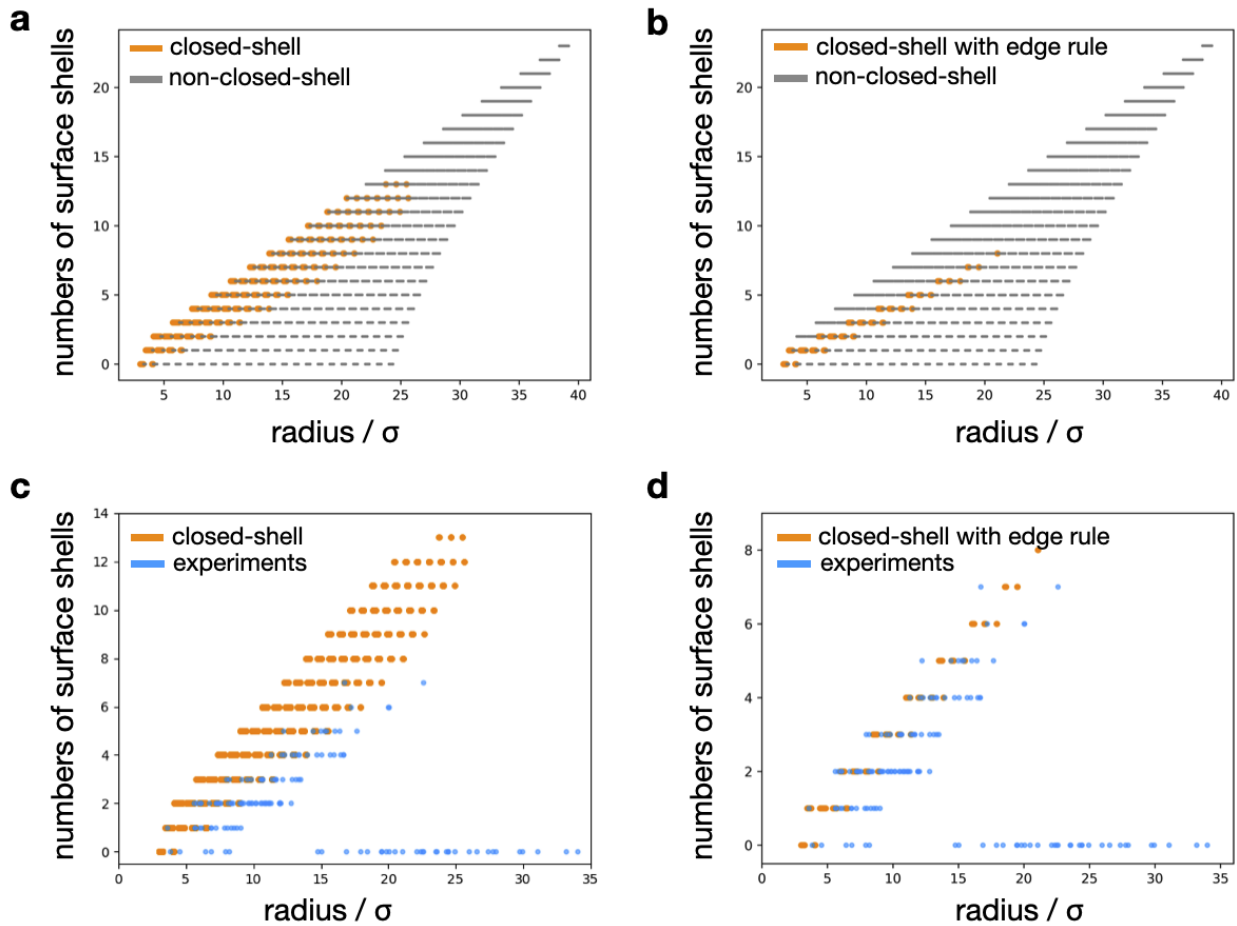

*Figure S13. Enumeration of icosahedral clusters with closed surface shells. a, The clusters with closed surface shell as a function of truncation radius based on the extended Mackay icosahedron with varying Mackay layers ($m = 3,4,...,23$). The orange region represents cluster radii, which allow shell closure. In the grey region, it is geometrically impossible to form closed-shell clusters. The number of Mackay shells, $m$, can be increased to infinity, which creates infinitely large icosahedral clusters and consequently increases the grey area. But the orange area does not increase anymore because closed surface shells cannot form. b, Applying the edge rule (Figures S9) selects a subset of possible closed-shell clusters and shift the failure of shell closure from a critical radius of about 25 to 20. Note that also without the edge rule, a breakdown of closed surface shells is predicted by the geometric model. c, d, Theoretical prediction (orange) compared with experimental observation (blue). The geometric model with the edge rule accurately predicts the breakdown of closed surface shells observed in experiment.*

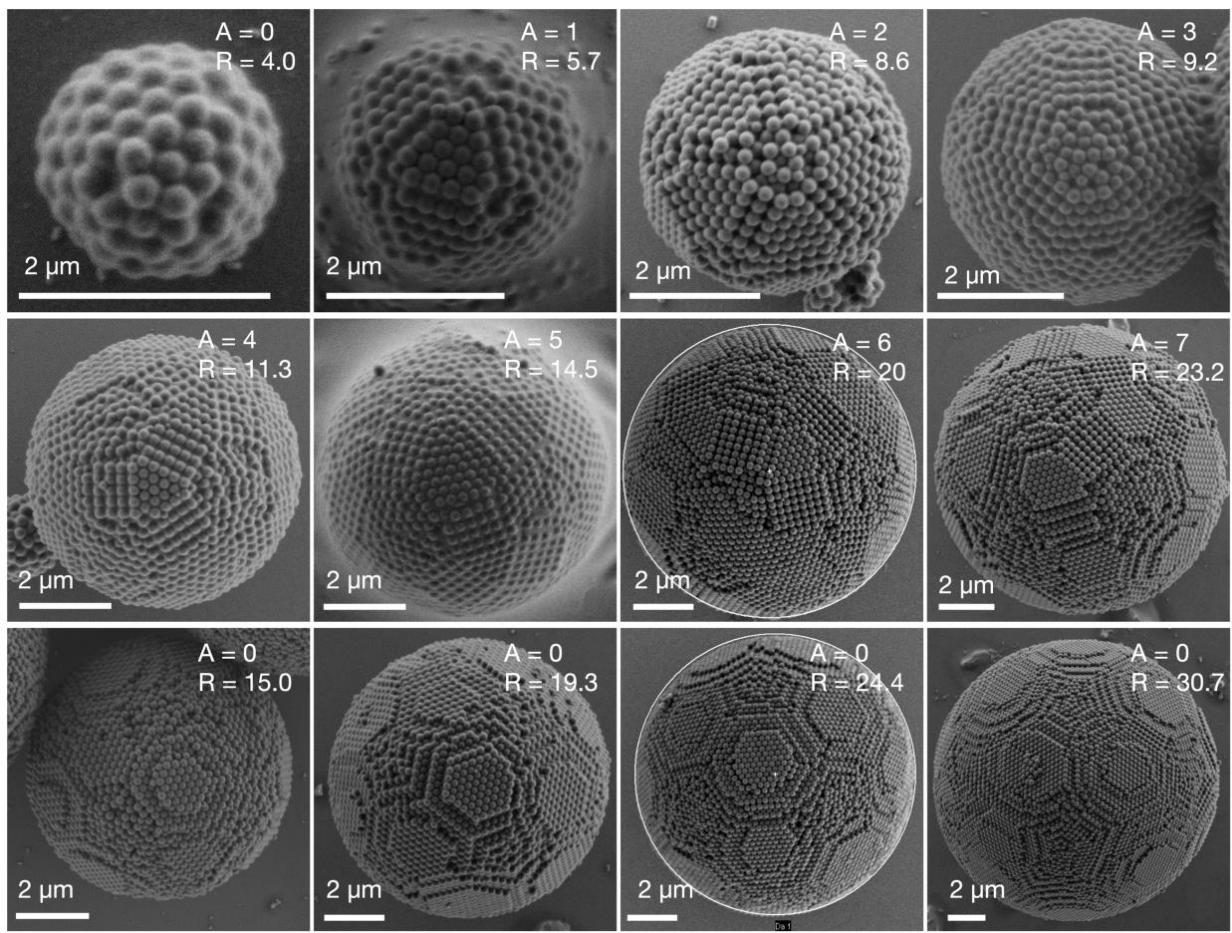

***Figure S14. Library of colloidal clusters of different sizes observed in SEM.*** *Top two rows are anti-Mackay clusters corresponding to magic number clusters with closed surface shells. Radius (R) are measured from the SEM images, number of anti-Mackay shells (A) are deduced from the width of the rectangular regions at the surface (Figure S1, S3). The bottom row shows non-closed-shell football cluster. Disconnected terraces can be seen on the cluster surface that breaks the closed surface shell ($A = 0$). For the transition region ($R > 15$, Figure 2a) both features, a region of closed anti-Mackay surface shell and a region of non-closed football surface shell can be simultaneously observed.*

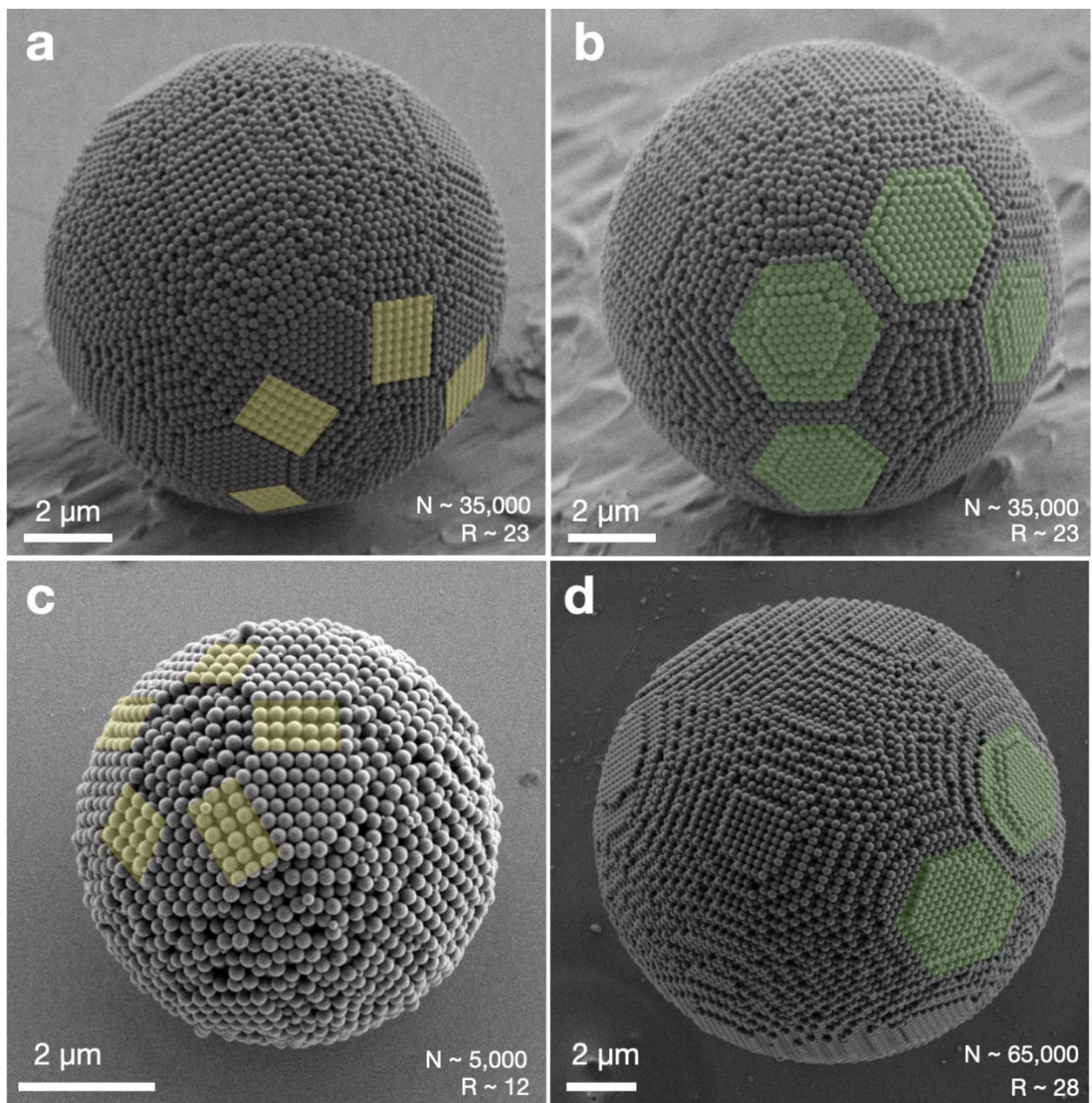

*Figure S15. Decahedral clusters observed in SEM. a, Anti-Mackay-type decahedral clusters whose surface around five-fold symmetric axis is tiled with connected {111} hexagon and {110} rectangles (marked in yellow). b, Football-type decahedral clusters, whose surface shows terraces of {111} planes. Decahedral clusters in (a,b) are in the coexistence region. c,d, Decahedral clusters in the anti-Mackay and the football occurrence region of the diagram (Figure 2A, Figure 5A).*